\documentclass{PoS}

\title{Cosmic Rays: What Gamma Rays Can Say}

\ShortTitle{Cosmic Rays: What Gamma Rays Can Say}

\author{\speaker{Roberto Aloisio}\\
        INAF - Osservatorio Astrofisico Arcetri, largo E. Fermi 5, I-50125 Firenze, Italy \\
        Gran Sasso Science Institute (INFN), viale F. Crispi 7, I-67100 L'Aquila, Italy\\
        E-mail: \email{aloisio@arcetri.astro.it}}

\abstract{ We will review the main channels of gamma ray emission due to the acceleration and propagation of cosmic rays, discussing the cases of both galactic and extra-galactic cosmic rays and their connection with gamma rays observations. }

\FullConference{Science with the New Generation of High Energy Gamma-ray experiments, 10th Workshop - Scineghe2014\\
		04-06 June 2014\\
		Lisbon - Portugal}
		
\begin{document}

\section{Introduction}

After more than 100 years from the discovery of Cosmic Rays (CR) the physics of this fascinating phenomenon was extensively studied, for a recent review see \cite{Blasi2013,Ptuskin2013} and references therein. In the energy range that spans from few GeV/n up to $10^3$ TeV/n a self-consistent scenario of the production and propagation of CR was developed in the last 30 years: the so-called standard model of galactic CR. This theoretical framework is based on two main pillars: (i) CR acceleration takes place in Super Nova Remnants (SNR) and (ii) CR propagation is diffusive in the Interstellar Medium (ISM). 

The idea that SNR are the sites of CR acceleration dates back to '30s \cite{Baade1934} and it has a twofold justification. From one side, SNRs are natural places in which strong shocks develop and such shocks can accelerate particles almost at the observed energies \cite{Bell1978,Blandford1978}. On the other side, SNRs can easily account for the required energetics \cite{Baade1934}. Nowadays, as a general remark, we can state that there is no doubt that galactic CR are accelerated in SNR, the open questions are which kind of SNR and which phase of the SNR evolution really do accelerates particles \cite{Blasi2013}.

At the highest energies in the regime of Ultra High Energy Cosmic Rays (UHECR), with energies $E>10^{17}$ eV, the theoretical framework aiming at the explanation of the observations is less refined respect to the lowest energies case. These particles are certainly of extragalactic origin \cite{Aloisio2012} and their propagation features trough astrophysical backgrounds seems well understood \cite{Aloisio}. Nevertheless, their possible sources and production mechanisms are still enshrouded in mystery. 

In the present paper we will discuss the observation of secondary $\gamma$-rays produced by the interaction of CR with the surrounding medium. We will consider the case of galactic CRs, focusing on the observations from SNRs and Molecular Clouds (MCs), and the case of extragalactic CR, focusing on the observations of the diffuse extragalactic $\gamma$-ray background and the $\gamma$-ray emissions from distant Blazars. We will show how these observations might unveil many unsolved problems in the physics of CR.

The observation of secondary emissions produced by CR while being accelerated in SNR is certainly one of the most powerful tools we can use to understand the physics of galactic CR acceleration. The observation of radio emissions by relativistic electrons, emitting through the synchrotron mechanism, started already in '50s, provides a first hint of the possible presence of amplified magnetic fields and accelerated particles in the SNR environment. However, the presence of relativistic nuclei, the actual smoking gun of CR acceleration, can be unambiguously proved only through the observation of $\gamma$-rays produced by the decay of neutral pions, product of the nuclear collisions between CR particles and the background plasma though the process $pp\to \pi^0\to\gamma\gamma$. This mechanism of $\gamma$-rays emission is called {\it hadronic}. Another mechanism, working in the same energy band $GeV-TeV$, that could be responsible for $\gamma$-rays emission in SNR is based on the Inverse Compton (IC) scattering of relativistic electrons on some photon background, this mechanism is called {\it leptonic}. Of course, if the latter mechanism is the ultimate responsible for the observed $\gamma$-rays emission it would not imply any confirmation or disproval of the SNR paradigm in the acceleration of CR. 

A very important step forward in this field of research was achieved in the recent years with an impressive amount of experimental data at TeV energies, by Cherenkov telescopes (HESS, MAGIC, VERITAS), and at GeV energies, by the Fermi-LAT and AGILE satellites. These experimental data are extremely useful to understand the origin (leptonic vs hadronic) of the observed $\gamma$-rays. The most important difference between the two alternative scenarios is related to the slope of the photons spectrum: the hadronic production gives the same slope of the ions distribution, while IC gives rise to a significantly flatter spectrum. In the regime in which synchrotron losses are negligible, ions and electrons spectra have the same behaviour ($\propto E^{-\gamma}$). The expected photon spectrum in the case of hadronic production is $\propto E^{-\gamma}$ while in the case of leptonic production will be $\propto E^{-(\gamma+1)/2}$.  

The emission of $\gamma$-rays from SNRs are connected not only with the actual acceleration region inside the remnant but can also dependent on the local environment surrounding the SNR. There are observations of $\gamma$-rays emissions from MC that, placed nearby a SNR, act as target for hadronic interactions giving rise to the pion production process \cite{AgileMC,FermiMC}. A SNR close to a MC represent a very interesting astrophysical system that can be studied in $\gamma$-rays not only as a diagnostic of CR acceleration but also as a laboratory to test CR escape from the source. The latter phenomenon being the actual CR injection in the ISM, therefore of paramount importance in the study of CR propagation \cite{Gabici2009}. 

Together with the emission from MC nearby a SNR, there are interesting systems in which isolated MC were observed as $\gamma$-rays emitters. This case is of particular importance in the study of the diffusive propagation of CR, offering the unique possibility of determining the CR spectrum unaffected by local effects such as the solar modulation. An interesting instance of these systems is represented by the $\gamma$-ray emission, already detected by by COS-B \cite{COS-B}, EGRET \cite{EGRET} and more recently by Fermi \cite{FermiGMC}, from the Gould Belt clouds, the nearest Giant Molecular Cloud (GMC). 

Gamma rays observations have an important impact also in the study of extragalactic cosmic rays, namely UHECR. These particles, with energies $E>10^{17}$ eV, that represent the highest energetic particles in the universe, depending on their chemical composition, interacting with astrophysical backgrounds, such as the Cosmic Microwave Background (CMB) and the Extragalactic Background Light (EBL), give rise to the process of pair production ($p\gamma\to e^{\pm}$) and photo-pion production ($p\gamma\to \pi^{0,\pm}$) that, in turn, produce electromagnetic cascades with the emission of a diffuse extragalactic $\gamma$ radiation in the energy band $10^{-2} \div 10^2$ GeV. Therefore, the data on the diffuse extragalactic $\gamma$ background can be extremely useful to constrain source models of UHECR \cite{Berezinsky2011,Gelmini2013,Roulet2013}. The observation of $\gamma$-rays from distant blazars is another important tool that could unveil the origin of UHECR. As was shown first in \cite{Blasi2005} and more recently in \cite{Essey2011}, the observed high-energy $\gamma$-ray signal may be dominated by secondary gamma rays produced along the line of sight by the interaction of UHECR with background photons. This possibility would explain the surprisingly low attenuation of $\gamma$-rays observed by distant blazars, because the production of secondary $\gamma$-rays occurs, on average, much closer to earth than the blazar distance. It is important to emphasise the central role that UHECR chemical composition plays in the emission of secondary $\gamma$-rays. In the case of heavy nuclei $\gamma$ emission is strongly suppressed because the process of photo-pion production becomes subdominant respect to the process of photo-disintegration. In this particular case it is expected a substantial suppression of the emission in $\gamma$-rays by UHECR.

The paper is organised as follows. In section \ref{SNR} we will discuss the $\gamma$-ray emissions from SNR. In section \ref{MC} we will discuss $\gamma$-ray observations from isolated MC and its consequences on the models of galactic CR propagation. In section \ref{UHECR} we will discuss $\gamma$-rays observation from distant blazars and the diffuse extragalactic $\gamma$-ray background. We will draw our conclusions in section \ref{Conclude}.

\section{Gamma Rays from Super Nova Remnant}
\label{SNR}

The success of the SNR paradigm for the acceleration of galactic CR relies on the first order Fermi mechanism that, using the test particle approximation, leads to a power law spectrum of the accelerated particles $N(E)\propto E^{-\gamma}$ with $\gamma=(r+2)/(r-1)$ being $r$ the shock compression ratio, namely the ratio between the velocity of the hotter shocked plasma and the velocity of the cold unperturbed one. In the case of a monoatomic gas with adiabatic index $\eta=5/3$, being $r=(\eta+1)/(\eta-1)$ for strong shocks\footnote{Shocks with sonic Mach number much larger than 1.}, one gets the very general result $N(E)\propto E^{-2}$ \cite{Blandford1978}. This result, of the so-called Diffusive Shock Acceleration (DSA), attains to the case of test particle approximation, nevertheless taking into account the DSA efficiency it is easy to see that the acceleration process channels a large fraction of the fluid ram pressure into CR. In this case CR particles cannot be considered as test-particles and their back-reaction on the shock has to be included in the theory, leading to the non linear theory of DSA (NLDSA) \cite{Blasi2013}. 

The most important consequence of non-linear effects in DSA is the formation of a shock precursor that slows down and compresses the upstream fluid. This fact produces a weaken of the proper shock and accelerated particles, diffusing in this modified fluid profile, experience different compression ratios depending on their energy. The resulting CR spectrum will no longer be a simple power law. Predictions of NLDSA, with CR back reaction on the shock, show a concave CR spectrum with a behaviour steeper than $E^{-2}$ at low energy and flatter at the highest energies. 

The hadronic emission of $\gamma$-rays, as discussed in the introduction, traces the CR spectrum in a unique way. Therefore, the observation of concave $\gamma$-ray emission from a SNR would represent the smoking gun for very efficient CR acceleration. Unfortunately the available $\gamma$-ray observations do not show any concavity in the spectra with spectral behaviours which are always steeper than $E^{-2}$ (at GeV energies and above) \cite{SNRCatalog}. This result seems at odds with the NLDSA expectations. Nevertheless, non-linear effects of CR are not limited to the shock hydrodynamics also involving the local magnetic field, which results amplified by the super-Alfvenic streaming of CR. In a coherent NLDSA picture CRs generate the magnetic field turbulence responsible for their own diffusion, this phenomenon has a central role in NLDSA and eventually allows particles to reach the observed energies (i.e. the knee energy around $10^{15}$ eV). The effect of magnetic field amplification is not only restricted to the maximum achievable energy, but it modifies also the compression factor $r$ because it implies fast moving scattering centres, with a velocity corresponding to a sizeable fraction of the shock speed \cite{Blasi2013}. In this way the compression factor, depending on the helicity of the waves upstream, can be decreased resulting in steeper spectra of the accelerated particles \cite{Blasi2013,Morlino2012,Caprioli2011}. 
\begin{figure}[htb]
    \begin{center}
        \includegraphics[scale=0.43]{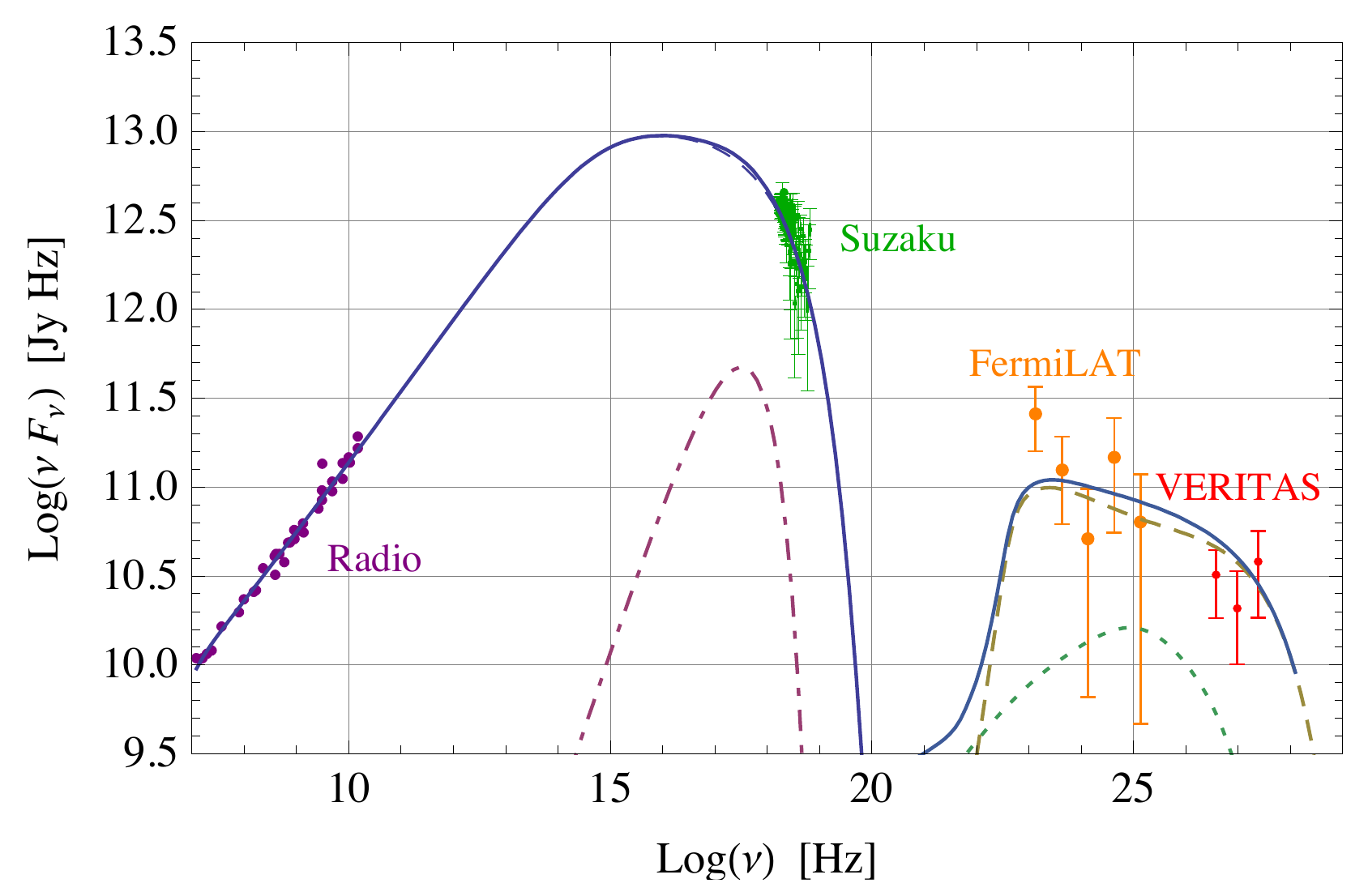}
        \includegraphics[scale=0.76]{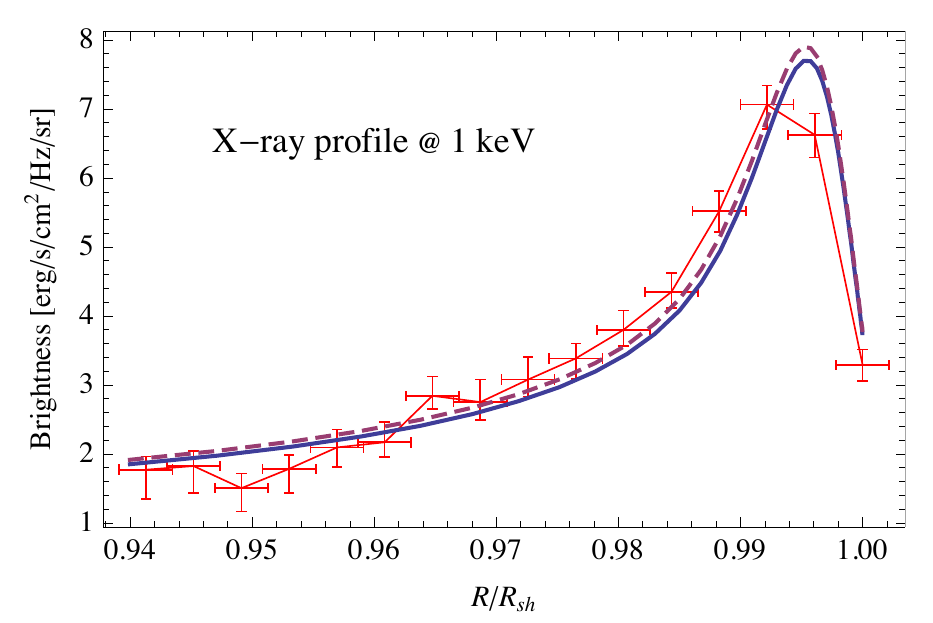}
        \caption{\it [Left Panel] Spatially integrated spectral energy distribution of Tycho. The curves show synchrotron emission, thermal electron bremsstrahlung and pion decay as computed in \cite{Morlino2012}. Gamma rays data from Fermi-LAT \cite{Giordano2012} and VERITAS \cite{Acciari2011} are shown. [Right Panel] Projected X-ray brightness at 1 KeV. Data points are taken from \cite{Cassam2007} the solid line are the computation of \cite{Morlino2012} after convolution with Chandra point spread function. (Both figures are taken from \cite{Morlino2012})}
\label{fig1}
    \end{center}
\end{figure}

The best opportunity to test theoretical models on the origin of galactic CR in SNRs is by considering the multifrequency emissions observed by these objects. Following the approach of \cite{Blasi2013} we will not list all SNR observed in $\gamma$-rays but we will discuss one specific case of SNR, that is sufficiently isolated to be considered as an individual source, discussing its $\gamma$-rays emission showing the type of information one can gather comparing theory and observations. 

The clearer case is that of the Tycho SNR, the leftover of a SN type Ia exploded in the homogeneous ISM as shown by the regular (spherical) structure of the remnant. Tycho shows a multifrequency spectrum that extends from the radio band up to $\gamma$-rays, with a thin X-ray rim observed all around the remnant with a spherical morphology. 

The observations of X-ray thin filaments, that is produced by synchrotron radiation of relativistic electrons at the shock, emitting radiation as $\nu(MHz)=3.7 B(\mu G) E^2 (GeV)$, confirms the presence of an amplified magnetic field of the order of $300 \mu G$. The spectrum of $\gamma$-rays observed by Fermi-LAT in the GeV region \cite{Giordano2012} and the spectrum of $\gamma$-rays observed by VERITAS in the TeV region \cite{Acciari2011} show a quite steep spectrum with a power law index at GeV energies of around $\alpha=2.3$ \cite{Giordano2012} and around $\alpha=2.0$ at TeV energies \cite{Acciari2011}. 

The two different sets of observations, in X-rays and $\gamma$-rays, can be accommodated fairly well in the framework of NLDSA. As shown in \cite{Morlino2012}, the steep hadronic spectrum comes from fast moving turbulent waves, CR scattering centres induced by the super-Alfvenic streaming of CRs themselves, with an Alfven velocity fixed by the amplified magnetic field. Therefore the $\gamma$-rays spectrum observed, with a clear hadronic origin, is directly linked to the strength of the amplified magnetic field, which, in turn, is the same quantity relevant to determine the observed X-ray morphology. Interestingly enough, still in the framework of a hadronic origin of the emission, the steep $\gamma$-ray spectra observed from Tycho can be explained also invoking environmental effects \cite{Berezhko2013}. 

The multifrequency emission of Tycho is shown in the left panel of figure \ref{fig1} and the X-ray brightness of its rims in right panel (both figures are taken from \cite{Morlino2012}).  The dash-dotted line in left panel shows the thermal emission from the gas downstream of the shock, assuming that the temperatures of electrons and protons in the plasma just behind the shock are related by $T_e=(m_e/m_p)T_p$ \cite{Morlino2012}. The short-dashed line in left panel represents the IC (leptonic) contribution to $\gamma$-rays emission and the dashed line the contribution of pion decay (hadronic) \cite{Morlino2012}. The solid line represents the total flux. It is rather impressive how the magnetic field required to explain the radio and X-ray emissions as synchrotron radiation also gives a compelling description of the thickness of the X-ray rims. 
\begin{figure}[htb]
    \begin{center}
        \includegraphics[scale=0.37]{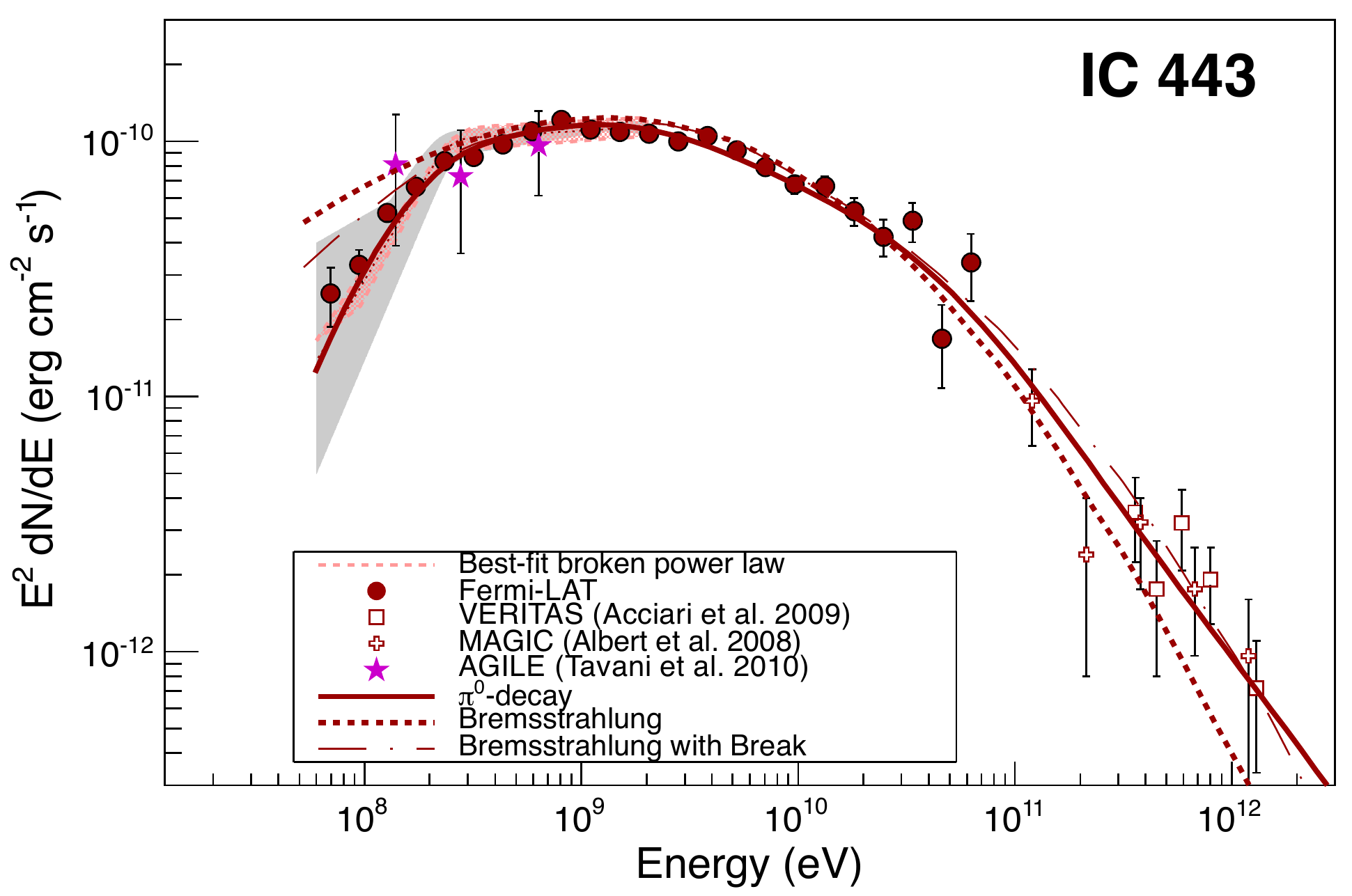}
        \includegraphics[scale=0.39]{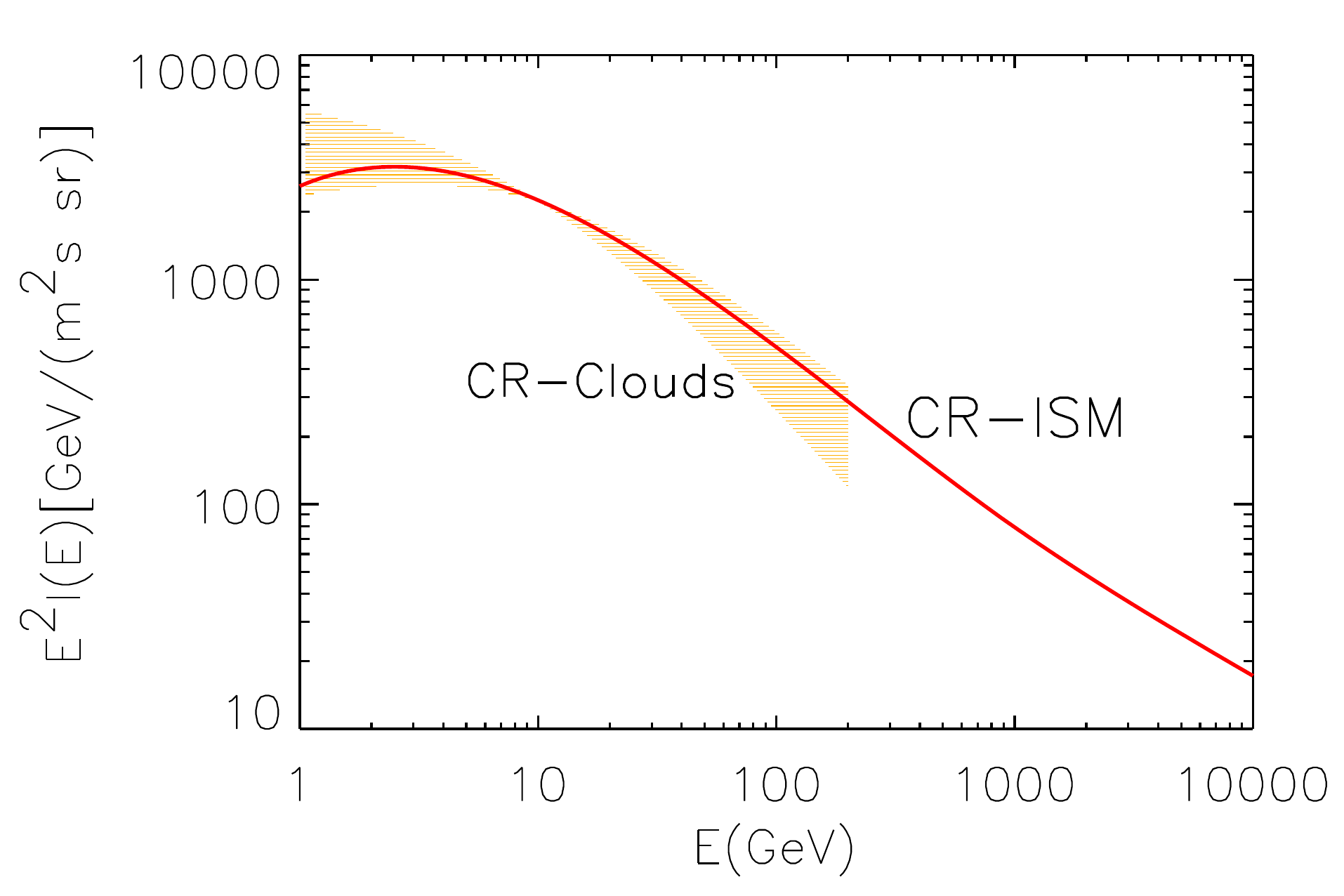}
        \caption{\it [Left Panel] Pion bump in the $\gamma$-ray emission of SNRs IC 443 (the plot is taken from \cite{FermiMC}). [Right Panel] Proton spectrum as computed in \cite{Aloisio2013} compared with the spectrum of CR (yellow shadowed region) as inferred from $\gamma$ ray observations of clouds as discussed in \cite{Neronov2012}.}
\label{fig2}
    \end{center}
\end{figure}
Assuming a Bohm diffusion regime at the shock, these results imply a maximum acceleration energy at the level of $E_{max}\simeq 500$ TeV and a compression ration at the shock that, according to \cite{Morlino2012}, can account for the steep $\gamma$-ray spectra observed. 

Let us conclude this section stating how the case of Tycho can be regarded as a workbench in which we can test the credibility of NLDSA models. In the near future the new generation $\gamma$ rays telescopes, most notably the Cherenkov Telescope Array (CTA), will allow the discovery of a considerable number of new galactic SNRs that are in the process of accelerating CR. The increased angular resolution of those telescopes will enable the measure of $\gamma$ rays from different regions inside the same remnant so to achieve a better discrimination of the link between the acceleration process and environmental effects.   

\section{Gamma Rays from Molecular Clouds}
\label{MC}

Recently two different collaborations Agile and Fermi-LAT \cite{AgileMC,FermiMC} claimed the detection of the most sought-after feature of the $\gamma$-rays spectrum: the so-called pion bump. A feature that directly links $\gamma$ rays production with CR propagation through the process $pp\to \pi^0\to \gamma\gamma$.

In the left panel of figure \ref{fig2} we show the $\gamma$-ray spectra observed by Fermi-LAT from the SNR IC443 were the pion bump is clearly recognisable \cite{FermiMC}. These observations were carried out looking at astrophysical systems in which a MC, with high gas density, seats nearby a SNR. These systems are particularly interesting, because the high density provides an increased probability (by a factor of about $10^{2}\div 10^{3}$) of $pp$ interactions. Therefore, observing the copious $\gamma$-ray emission produced one can gather important informations on CR propagation around the source and their escape process from it. Concerning the acceleration process itself, SNR with a MC nearby, being typically old objects, are not expected to still accelerate particles to very high energies \cite{Blasi2013}. 

The type of information one can actually get depends on the MC location respect to the SNR morphology. We can distinguish two different scenarios: one in which the shock is directly propagating inside the MC and one in which a MC, sufficiently faraway from the remnant, is illuminated by CR just escaped from the acceleration region. 

In the first case, being very small the fraction of ionised gas in the MC and very high the gas density itself, the collisionless shock wave propagating in the cloud may became collisional, completely changing the physics of the system that would result in the heating of the molecular gas. This instance seems confirmed by the observation of maser emissions from certain MC \cite{Herwit2009}.

The case of MC illuminated by CR already escaped the remnant and still in its proximity has a paramount importance in the determination of the actual process that injects CR in the ISM. For this reason this case has attracted more attention in the recent years (see \cite{Blasi2013} and references therein). In particular, the CR flux reaching MC is expected to be time dependent, it comes by both the time dependence of the escaping process and the diffusion time needed to CR particles to reach MC. If $R_{MC}$ is the distance separating MC and SNR, one can expect a low energy cut-off developing in the CR spectrum that corresponds to the energy at which the diffusion length matches $R_{MC}$: $\sqrt{D(E)\tau_{SNR}}\simeq R_{MC}$. A spectral break in the $\gamma$-ray emission from MC, linking this emission with CR propagation in the remnant proximity, might have been recently observed by the Agile satellite in the SNR W28 \cite{AgileMC}. These observations refer to the system of two clouds at different distances from the SNR that appear to be illuminated by different CR fluxes with a low energy spectral break placed at higher energies for the most distant cloud \cite{AgileMC}.  

Among $\gamma$ ray emissions from MCs also important is the case of isolated clouds, particularly those located appreciably above and below the galactic disc. These emissions are contributed mainly by $pp$ interactions and enable a direct determination of the CR flux penetrating the cloud, therefore the CR flux diffusing through the ISM. As recently pointed out in \cite{Aloisio2013,Neronov2012} from these studies it is possible to determine specific features in the CR spectrum related to their diffusive motion, independently of the local effects due to the Earth position in the solar system. 

As a general remark it should be pointed out that CR always play a crucial role in determining the diffusion properties of the medium in which they propagate. As discussed in the previous section, these non-linear effects are crucial in the acceleration sites, where they contribute to both the maximum achievable energy and the spectral shape of the accelerated particles \cite{Blasi2013}. Nevertheless, self generation of magnetic turbulence could play an important role also in the diffusive motion of CR through the ISM \cite{Aloisio2013}. 

Recently the PAMELA satellite observed a change of slope in the CR spectrum at energies around $200$ GeV \cite{PAMELA}, these observations\footnote{We point out here that the spectral break claimed by PAMELA, while it agrees with the highest energy observations of CREAM \cite{CREAM}, seems not confirmed by the recent AMS observations as shown at the last ICRC conference \cite{AMS}.} can be explained in terms of self-generated magnetic turbulence, that determines the actual CR diffusion coefficient at energies below the spectral break observed \cite{Aloisio2013}. The observations of $\gamma$ rays from isolated MCs indirectly confirmed that the spectrum of CRs with energy $10\lesssim E \lesssim 200$ GeV may be steeper than previously thought, and with a slope compatible with the one quoted by PAMELA in the same energy region \cite{Neronov2012}. These results are based on the analysis of the gamma ray emission detected by the Fermi-LAT from selected clouds in the Gould's belt.
In the right panel of figure \ref{fig2}, following the computations of \cite{Aloisio2013}, we plot the proton spectrum determined self-consistently taking into account particles diffusion on self-generated turbulence (continuos line) and the proton spectrum as inferred from $\gamma$-rays observations (yellow shadowed region) \cite{Neronov2012}. 
\begin{figure}[htb]
    \begin{center}
        \includegraphics[scale=0.365]{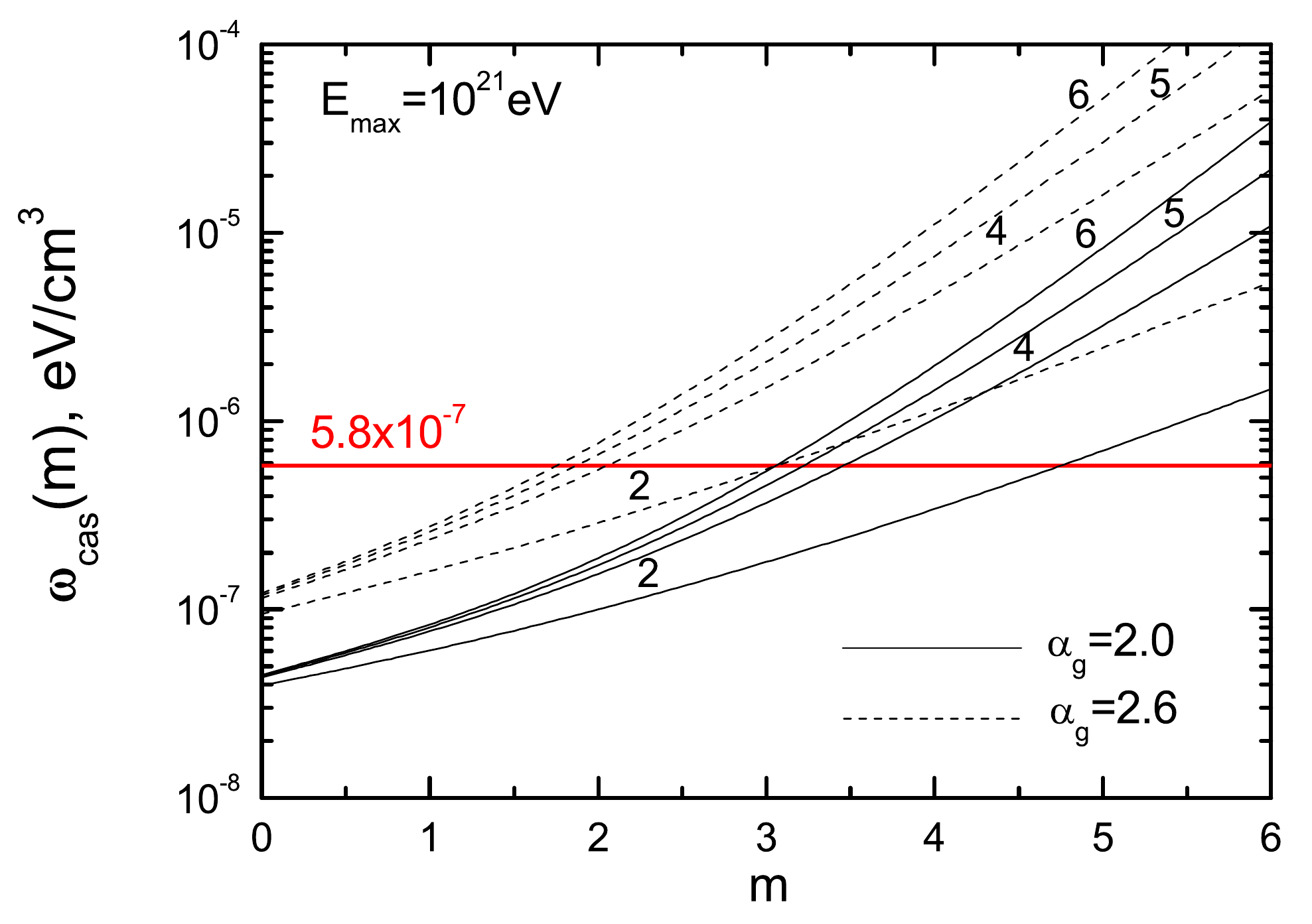}
        \includegraphics[scale=0.38]{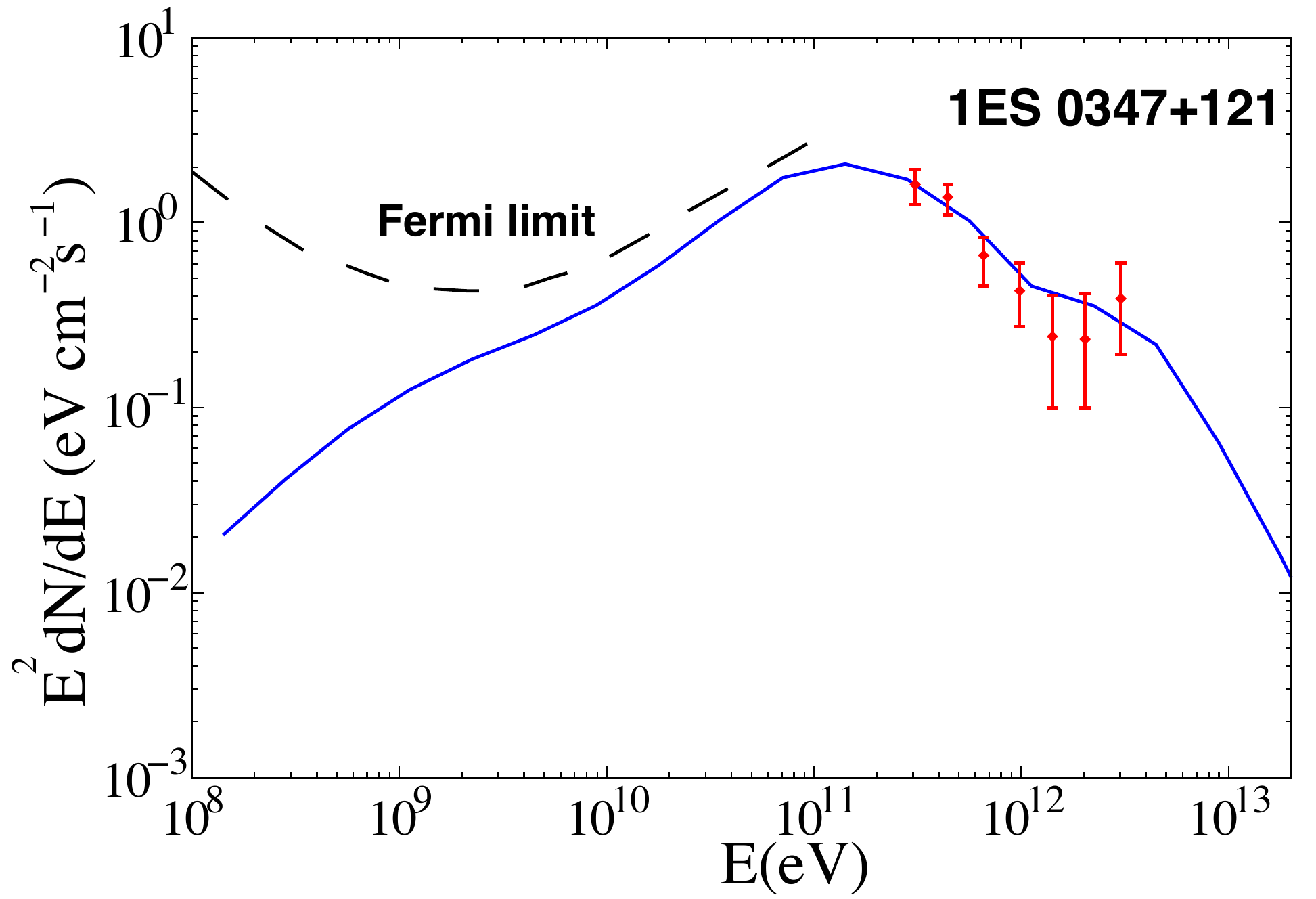}
        \caption{\it [Left Panel] Cascade energy density as function of the cosmological evolution of sources for different choices of the injection power law index (the figure is taken from \cite{Berezinsky2011}). [Right Panel] $\gamma$ rays observed by HESS from the Blazar 1ES 0347+121 compared with the theoretical flux determined from the interactions of UHECR along the line of sight (the figure is taken from \cite{Essey2011}).  
         }
\label{fig3}
    \end{center}
\end{figure}
From figure \ref{fig2} it is evident the fairly good agreement between observations and theoretical expectations. This result, still conditioned by large experimental uncertainties, shows the importance of $\gamma$ rays observed by isolated MCs that could unveil important details of the physics of CR propagation. 

\section{Gamma Rays from Ultra High Energy Cosmic Rays}
\label{UHECR}

Extragalactic CR with energies higher than $10^{17}$ eV, the so-called UHECR, propagating through CMB and EBL, give rise to the processes of pair-production and photo-pion production \cite{Aloisio}. These processes start electromagnetic (EM) cascades that end-up in the production of neutrinos and $\gamma$-rays. Depending on the chemical composition of UHECR, the amount of $\gamma$-rays produced in the cascades can represent a relevant fraction of the extragalactic diffuse $\gamma$-ray background observed by Fermi-LAT. Therefore, through these observations it is possible to constrain the characteristics of UHECR sources \cite{Berezinsky2011}, namely their possible cosmological evolution and spectrum. 

Following the results of \cite{Berezinsky2011}, assuming that UHECR are pure protons injected with a power law spectrum $\propto E^{-\alpha}$ by sources with cosmological evolution of the type $(1+z)^m$ ($z$ is the source red-shift and $m$ the parameter that drives the evolution), in the right panel of figure \ref{fig3} we plot the cascade energy density $\omega_{cas}$ as function of the sources cosmological evolution ($m$ parameter) for different choices of the injection power law index $\alpha$. The energy density of the $\gamma$-ray background measured by Fermi-LAT is shown through the red continuos line, from this figure it is evident the strong potential of these measurements that already enable the exclusion of strong cosmological evolution of sources, substantially restricting the astrophysical objects eligible for being UHECR sources. Let us conclude this part stressing the importance of UHECR chemical composition, if one releases the hypothesis of a pure proton composition, assuming the heavy composition claimed by Auger at the highest energies, it results in a substantial reduction of EM cascades and the flux of secondary $\gamma$ rays could easily fall below the detection threshold. 

Among possible UHECR sources a particular role might be played by Active Galactic Nuclei (AGN), as they show the right energetics and density expected on the basis of UHECR propagation studies \cite{Aloisio}. An interesting instance of the possible UHECR acceleration in distant Blazars was putted forward by \cite{Blasi2005,Essey2011}. The high energy $\gamma$ ray signal observed by those objects might be dominated by secondary $\gamma$ rays produced, along the line of sight, in the EM cascades triggered by the interaction of UHECR protons with astrophysical backgrounds. This hypothesis explains the surprisingly low attenuation of the $\gamma$ radiation observed, because the production of secondary $\gamma$ occurs, on average, much closer to the Earth respect to the distance of the source \cite{Essey2011}. The shape of the $\gamma$-ray spectrum is quite independent of the UHECR spectrum, while it depends mostly on the astrophysical backgrounds, i.e. CMB and/or EBL. The actual background that dominates the interactions depends on the maximum energy of UHECR: if $E_{max}>10^{19}$ eV CMB dominates, otherwise also EBL plays a role. 

In right panel of figure \ref{fig3}, just as an example, we plot the case of the Blazar 1ES 0347+121, observed by Hess, together with the spectrum $\gamma$-rays produced by EM cascades started by UHECR along the line of sight (the figure is taken from \cite{Essey2011}). The quite good agreement obtained shows the importance of this explanation of the observed TeV emission from distant blazars. Nevertheless it should be pointed out that this possibility critically depends on UHECR chemical composition and the Intergalactic Magnetic Field (IMF) strength. As before, in the case of nuclei the suppressed probability of EM cascades reduces the flux of secondary $\gamma$-rays below the detection thresholds. Moreover, for IMF larger than $3\times 10^{-14}$ G the spread of the EM cascade induced by the magnetic field prevents any observation along the line of sight \cite{Essey2011}.

\section{Conclusions} 
\label{Conclude}

We conclude just by stating the paramount importance that $\gamma$-ray observations play in the physics of cosmic rays. As it was discussed in the present paper these two signal carriers are intimately linked and, in a multi-messenger approach to high energy Astrophysics, should be always studied in connection. 

The importance of $\gamma$-ray observations is spread in all fields of CR physics. In the case of galactic CR the observation of $\gamma$-rays from SNR and MC can unveil the details of the acceleration and propagation processes. While in the case of UHECR the possible role of $\gamma$-ray observations depends on the chemical composition of those particles. In the case of protons, the copious production of secondary $\gamma$-rays due to the interaction with astrophysical backgrounds opens up to the possibility of testing different source models, in terms of both injection spectrum and cosmological evolution of the sources. 

The next generation $\gamma$-rays telescopes, from both sides of on ground and satellite detection, will start taking data in a few years bringing an unprecedented quality and quantity of new observations that will surely have a groundbreaking impact on the physics of cosmic rays.

\section*{Acknowledgements}
\noindent The author warmly thanks V. Berezinsky, P. Blasi and the whole High Energy Astrophysics Group of the Arcetri Astrophysical Observatory for joint work and continuos discussions on the physics of Cosmic Rays and related subjects. Giovanni Morlino is also thanked for a critical reading of the manuscript.

\end{document}